# A Survey on Crowdsourcing Applications in Smart Cities

Hamed Vahdat-Nejad*, Tahereh Tamadon, Fatemeh Salmani, Zeynab kiani-Zadegan, Sajedeh Abbasi, Fateme-Sadat Seyyedi

PerLab, Faculty of Electrical and Computer Engineering, University of Birjand, Iran

{vahdatnejad, tahereh.tamadon, salmani_fatemeh98, zeynab_kiani.zadegan2021, sajedeh_abbasi, fateme_seyyedi1997}@birjand.ac.ir

**Abstract**

With the emergence of the Internet of things (IoT), human life is now progressing towards smartification faster than ever before. Thus, smart cities become automated in different aspects such as business, education, economy, medicine, and urban areas. Since smartification requires a variety of dynamic information in different urban dimensions, mobile crowdsourcing has gained importance in smart cities. This chapter systematically reviews the related applications of smart cities that use mobile crowdsourcing for data acquisition. For this purpose, the applications are classified as environmental, urban life, and transportation categories and then investigated in detail. This survey helps in understanding the current situation of smart cities from the viewpoint of crowdsourcing and discusses the future research directions in this field.

## 1. Introduction

Technological developments have now turned human life objects into smart objects that can perceive and react to their surrounding environments [1]. These smart objects constitute the main part of the Internet of Things (IoT) [2]. With its smart solutions, the IoT has significant effects on all dimensions of human life [3]. It is used in different areas such as healthcare ADDIN EN.CITE [4-6], environment [7, 8], transportation ADDIN EN.CITE [9-11], security [12, 13], entertainment [14], business [15], and tourism [16]. After the previous three industrial revolutions of mechanization, electricity, and IT, the IoT and its relevant services are known as the fourth industrial revolution [17, 18]. According to previous studies, more than 50 billion devices have been connected to the Internet in 2021 [19] due to the pervasiveness of the IoT and its countless applications. By 2025, more than 100 billion devices will be connected to the Internet [20]. These figures indicate the ever-increasing growth of the IoT.

As the world's population grows, there has been an unprecedented upward trend in urbanization [21]; therefore, urban infrastructure has been under tremendous pressure. As a result, urban managers must modernize urban life, formulate effective strategies, and make initiative plans [22]. Therefore, different countries have made some efforts in urban services and infrastructure to develop cities with better facilities and improve socioeconomic conditions [23]. Their efforts lead to the development of smart cities, which can greatly help states solve socio-economic crises of the recent years [24]. Thanks to the IoT, smart cities are now growing rapidly [25, 26]. Different projects in smart cities have been conducted such as smart lighting [27, 28], smart parking [29, 30], smart agriculture [31, 32], smart waste management [33, 34], smart buildings [35, 36], and smart tourism [37, 38].

With the development of advanced mobile devices, crowdsourcing has emerged for the supply of information in smart cities to employ the potential capacity of citizens (crowds) for different tasks. Accordingly, crowdsourcing is an online process of collecting information through the participation of citizens via their smartphones [39, 40]. In fact, crowdsourcing is a participatory online activity in which an individual, an institution, an organization, or a private company recommends participation in an activity to a group of heterogeneous individuals. Users (crowds) participate in that activity by bringing their work, money, knowledge, and experience [41]. Collecting information in a smart city by crowdsourcing helps to obtain social feedback on a subject or offer solutions to a problem [42, 43]. Crowdsourcing has now attracted a great deal of attention both academically [44, 45] and commercially[1] [46].

Various survey studies have analyzed smart cities from different perspectives such as concepts and applications  ADDIN EN.CITE [25, 47-49], data management [50], security and privacy [51], and cloud computing [52]. Besides, many other survey studies have analyzed crowdsourcing in different aspects such as overviews [53], obstacles and barriers [54], incentive and task assignment [55], software engineering [56], security, privacy and trust [57], quality control [58], context awareness [59], and spatial crowdsourcing [60, 61]. However, investigating crowdsourcing in smart cities has specific characteristics that have been mostly neglected by previous survey papers. In this regard, two survey papers [62] [63] have recently investigated crowdsourcing in smart cities. The main distinguishing characteristic of the current chapter is the concentration on the application as it is the main perception of citizens from smart cities. Therefore, this review study collects, analyzes, and classifies the research applications that employed crowdsourcing in smart cities. To this end we used keywords of "crowd sensing" or "crowd sourcing" and "smart city" or "smart cities" in Google Scholar as the most comprehensive paper indexing service. We then refined the retrieved results in two steps: (1) We eliminated the papers that have been published by unknown or unpopular publishers and (2) We semantically checked the papers to eliminate those that mistakenly have been retrieved. After a review, these applications and research studies are classified into environmental, urban life, and transportation categories (table 1), each of which is discussed separately, in detail. This can help clearly understand the status quo as well as future research paths.

---

[1] https://www.upwork.com; https://www.mturk.com

This chapter consists of the following sections. Section 2 reviews the environmental applications of crowdsourcing in smart cities. Similarly, Section 3 and Section 4 investigate applications of urban life and transportation, respectively. In the end, Section 5 draws a conclusion and presents the future research paths.

**Table 1.** Research studies reviewed

| Research applications | Environmental | Urban life | Transportation |
|---|---|---|---|
| [64] | Flood forecasting | | |
| Mycoast [65] | Flood forecasting | | |
| Environmental Sensing [66] | Air pollution measurement | | |
| Airsense [67, 68] | Air pollution measurement | | |
| City Soundscape [69] | Noise estimation | | |
| UserVoice [70] | | Acquiring innovative ideas | |
| Publicsense [71] | | Reporting problems of public facilities | |
| [72] | | Reporting problems of public facilities | |
| CUAPS [73] | | Urban anomaly prediction | |
| mPASS [74] | | Providing personalized path for special users | |
| FlierMeet [75] | | Advertisement sharing | |
| [76, 77] | | | Acquiring Parking spaces |
| [78] | | | Acquiring Parking spaces |
| [79] | | | Acquiring Parking spaces |
| [80] | | | Acquiring Parking spaces |
| [81] | | | Smart public transportation |
| CrowdOut [82] | | | Road Safety |
| [83] | | | Fuel prices |
| [84] | | | Navigation |
| [85] | | | Music recommendation |

## 2. Environmental

The development of cities and human interference in nature have caused many problems such as floods, air pollution, and noise pollution. It is essential to monitor the environment, especially the urban environment, because it can raise public awareness of human living environments. In this regard, an efficient solution to environmental monitoring is mobile crowdsourcing. For this purpose, mobile users are required to collect and report their problems in an urban environment [86]. To collect information through crowdsourcing, it is necessary to prepare the platform consisting of user mobiles and urban servers [87]. Urban managers are then informed about environmental problems in time so that the necessary actions could be taken to solve the problems.

Floods are among the natural disasters that can cause great loss of life and financial damage. For many years, human societies have been trying to reduce the consequent costs and damages by predicting the occurrence of floods. Unfortunately, many cities have no flood warning systems. With the help of crowdsourcing, early warning systems can predict floods [88]. For this purpose, an Android application and a Web-based system have been presented to predict floods through crowdsourcing [64]. For participation, users must have smartphones with specialized sensors such as accelerometer, gyroscope, magnetometer, barometer, inertial measurement unit (IMU), and GPS. When users take photos of the edge of the water surface (*e.g.* river banks), data of mobile sensors are processed through a geometric method (*i.e.* 3-cone intersection method) to determine the altitude and pitch angle. The resultant information is then sent to servers along with the photos. Finally, a map of information on floods and rivers will be shown to users and urban managers via a Web application. In MyCoast [65], experts take photos of districts in which they think floods are likely to occur and provide their comments containing estimation of water depth and reduction or rise level of water. The data is then sent with spatiotemporal information to the server where convolutional neural networks are employed to process the images for the prediction of flood occurrence. Moreover, the tweets that have flood-related keywords such as flood, inundation, dam, dike, and levee are extracted. The relevant tweets are then processed through natural language processing (NLP) methods to extract their spatial information. Based on crowdsourcing and Twitter data , the server finally predicts the flood location and presents it on Google Map.

Air pollution is among the most prevalent problems in metropolises where the citizens need to know about the quality of air in their neighborhoods [89]. In environmental sensing [66], citizens can measure the concentration of environmental pollutants such as carbon monoxide, nitrogen dioxide, temperature, and humidity by using mobile environmental sensors (*e.g.* battery, onboard microprocessor, firmware, and Bluetooth) that they carry while walking or cycling. This information is then sent to a unified sensing platform and is processed through different analysis algorithms. If the measured environmental data of a district is incomplete, data-driven regression and interpolation techniques are utilized to complete the data. Ultimately, the air pollution information is presented as an updated heat map to users. In AirSense [67, 68], air quality monitoring devices (AQMD) send their IDs and data of measured air quality via Bluetooth to the nearest smartphone on which the program is installed. The received data is cleaned and formatted and then sent to the cloud. In the cloud, the OpenShift [90] service provider (*i.e.* the provider of free platform-as-a-service that provides a platform to store, aggregate, and analyze information obtained from sensors) is employed to collect, aggregate, analyze, and store data. Finally, at the request of users, the local air quality index as well as the air quality index map (AQI map) are sent to them with regard to the neighborhood in which they are located.

Noise pollution is another problem that metropolises face as their populations grow. It is so serious that it can have various physical and mental effects on citizens and jeopardize their health. Hence, states need certain information and sources to offer specific solutions to this problem [91]. City Soundscape [69] is a platform presented through crowdsourcing for realizing large-scale and low-cost acoustic measurement. Utilizing this application, users can record noise both automatically (when the application collects data automatically) and manually (when users set the starting point and duration for noise measurement) through the built-in mobile microphones. They can then send the recorded noise to the server. Users can also make comments or take a photo of a place where there is noise and represent their dissatisfaction with the environment by providing a numerical psychometric scale to help the perception of noise pollution levels. Data processing is then performed on the server in accordance with the extract-transform-load (ETL) pipeline (*i.e.* a process by which information is collected and processed through one or several different sources and then uploaded in the database). Ultimately, objective noise maps are created along with subjective noise maps, which are based on user comments.

## 3. Urban Life

With scientific advances at the dawn of the Industrial Revolution, many changes occurred in urban structures, resulting in modern urbanization. Gradually, with population growth, urban infrastructure has sustained a great deal of pressure; therefore, managers developed cities based on smart city frameworks [22]. Information technology (IT) now plays a key role in the smartification of cities [92], for which mobile crowdsourcing can collect a plethora of information from cities to help states improve the lives of citizens [22]. In this section, we review the studies that used mobile crowdsourcing for urban life smartification.

Urban managers need innovative ideas to plan urban projects for the smartification of cities. UserVoice [70] offers an online interactive platform for the collection of citizens' ideas through mobile crowdsourcing. In addition to recording their ideas, citizens can vote and comment on the ideas of others. Every user is allowed to vote on three ideas. To encourage the users, an Apple iPad2 was awarded to one user at random. Experts then evaluate the ideas based on three criteria: innovation, feasibility, and benefit for users by using Krippendorff's alpha (*i.e.* an index that shows the extent agreed among experts on the evaluation of quality parameters). The mean of every quality dimension of data is then calculated to select the ideas with the highest ranks for the urban development stage.

The failure of urban facilities can affect the quality of life among citizens. Therefore, a crowdsourcing platform has been developed to report the problems of public facilities in smart cities [71]. Users enter their name, phone number, type of damage, and corresponding photos to complete a report on the application and send it to the server. After normalizing these photos, the server classifies the data sent by users. Finally, users can look up different reports on a city map and become aware of details of a problem, responses, and recent events. Since many problems are reported by many users, it is important to identify repetitive reports. For this purpose, repetitiveness or genuineness of a report is determined with respect to temporal dimensions, two spatial dimensions (*i.e.* latitude and longitude), and different categories of reports (*e.g.* street light, traffic sign, and pothole) as well as the least-squares distance and the Bayes theorem [72].

In metropolises, people of different opinions and cultures live together, something which can cause some citizens to show certain behaviors or experience abnormalities. CUAPS [73] has been proposed for the timely prediction of urban abnormalities (*e.g.* noise, illegal use of public facilities, and urban infrastructure malfunctions). Citizens can send their complaints to this system. The city districts are then clustered through a Bayesian inference model based on the number of abnormalities and their temporal vectors (showing when abnormalities occurred). Finally, the Markov trajectory estimation is employed to predict the abnormality of every district based on the dependency of districts and a history of abnormality in that district. The timely prediction of abnormalities can help prevent them and bring peace to citizens.

Another problem with urbanization is the negligence of people with special needs (*e.g.* children, the elderly, and the disabled), for the commuting of whom the urban facilities are not usually appropriate [74]. With the help of users, mPASS [93] collects information on facilities/barriers such as stairs and ramps across a city. After users determine origins and destinations on the application, a recommended series of personal paths for walking (based on facilities/barriers) and traveling by bus (with the required equipment for citizens of special needs) will then be shown to users.

Advertising has a major role in urban life. FlierMeet [75] offers an advertisement sharing system through crowdsourcing. Utilizing their mobile cameras, users take photos of notice boards across the city and send them to the server via this application. GPS, light sensor, accelerometer, and magnetometer of smartphones are also employed to measure and send the position, light intensity, motion blur, and shooting angle. The server evaluates the quality of photos and deletes the low-quality or repetitive ones. The contents of fliers are also labeled with tags including the advertisement, academic event, notice, and recruitment. Based on people's interests, the fliers are then labeled with semantic tags including popular (*i.e.* the fliers that are used by most people and reported from different locations), hot (*i.e.* the fliers that have extensive audiences and become extremely popular in a short time), professional (*i.e.* the fliers that are considered for a specific population with common needs and skills), and surprising (*i.e.* the surprising ads). Users can see their favorite fliers on the map, based on the classification or semantic tags and also make comments on the fliers.

## 4. Transportation

The increasing rate of transportation in smart cities has caused various issues such as traffic, accidents, and insufficient parking spaces on streets. Therefore, urban transportation systems are known as an important aspect in modern urbanization, becoming smart with technological advances. Mobile crowdsourcing is a paradigm for the real-time collection of a great deal of information on smart transportation in cities [94]. Based on crowdsourcing, the intelligent transportation system (ITSs) can identify the states of roads and offer quick and safe routes to citizens by predicting traffic across the city [95]. These systems can also provide other services such as crowdsourced geospatial data acquisition, urban traffic planning and management, smart parking, and green transportation [96]. This section reviews the papers that have used mobile crowdsourcing in different areas of urban transportation.

Finding a parking spot has now become a challenge to drivers at rush hours in metropolises. Mobile crowdsourcing has been employed to propose a smart parking system [76, 77]. Drivers can manually record the number of parking spots on the street on a questionnaire and then send it. The server first assumes that all parking spots on-street sides are unknown. After data of user questionnaires are received, the status of every spot is changed to occupied or available. Finally, available parking spots are offered to users in graphics or voice. Users are also provided with other pieces of information such as parking price, statistics about the arrival rate of vehicles, and parking rates.

A mobile crowdsourcing application [78] has also been proposed to show the map of available parking spots on the street. To employ this application, drivers must attach their smartphones vertically to their windshields when their magnetometer, gyroscope, and GPS sensors are activated. The orientation of axes on the magnetometer is important while processing its sensor signal. Using the information received from smartphones, the server infers the status around the user's car in terms of parking space. Another study [79] leverages a vehicle equipped with a sonar sensor, ultrasonic rangefinder, GPS, camera, Raspberry Pi chip, 3G/4G antenna, and 3G/4G connection. The ultrasonic rangefinder sensor creates a short pulse (in every 50 milliseconds) to measure the distance between a car and the roadside. Vehicles are then distinguished from the roadside barriers through the features extracted from the sonar sensor including the vehicle length, the distance to a parked car, the standard deviation of the distance, and the angle between vertices as well as the bottom of the detected object. After collecting and aggregating the parking spot information (*e.g.* parking capacity and location), the server generates parking capacity and spots on a map.

Developing technology-based transportation projects can impose hefty costs on organizations and companies; therefore, it is necessary to analyze a project before it is implemented and operationalized in order to identify and solve the potential issues. For this purpose, a crowdsourcing simulator has been implemented in Java to extract urban parking spots [80]. This simulator uses MASON [97] as a multi-agent simulation toolkit and utilizes Mapzen[2] to download geographic data. With this simulator, users can report occupied or available parking spots to the server. Moreover, the aging algorithm is adopted to estimate the validity of a parking spot (*i.e.* a report will be unreliable after two minutes, whereas it will be invalid and then deleted from the system after five minutes). Finally, OpenStreetMap (OSM) [98] services are used to display the map of available parking spaces obtained from users' participation.

The public transportation system now plays a significant role in moving citizens. People are willing to use the public buses for various reasons such as low costs and ease of use. The challenge is the lack of awareness regarding the arrival time of buses at stops that wastes people's time. In this regard, a mobile crowdsourcing application [81] with the help of the Beacon[3] technology has been presented to provide awareness of the position of buses as well as the approximate time of their arrival at different stops. The transmitter of the Bluetooth-based beacons installed on buses and bus stops broadcasts certain signals with its ID to the surrounding environment. A user's smartphone receives the bus stop ID and sends it to the server to inquire about the arrival times of buses. Upon the arrival of a bus boarding the user, a message containing information on the bus and the bus stop is sent to the server. The current position of the bus is also sent periodically to the server. Finally, the server estimates the arrival time of a bus at the next stop and notifies citizens so that they can use the public transportation system with appropriate timing.

In metropolises, non-compliance with traffic laws and road defects can cause issues. CrowdOut [82], a crowdsourcing-based service, provides citizens with the ability to report traffic violations and road defects. If users witness violations (*e.g.* unauthorized speed and unauthorized parking) and defects (*e.g.* traffic light failure, damaged roads, and traffic), they can report the problem. They can also take photos, add provide short comments about the type of violation (*e.g.* parking violation) with spatial coordinates (GPS), and send them to the server. After aggregating the received data, the server can display a map of problems and violations to urban managers via Google Map. Finally, urban managers can also observe the photos as well as various diagrams of violations in different districts to take the necessary actions (*e.g.* contacting the owner of the vehicle and alerting the impound service to remove the vehicle).

---

[2] https://www.mapzen.com/
[3] Https://developer.apple.com/ibeacon/

Given the fuel price differences at different gas stations in some countries, drivers are willing to fuel their cars at the stations that offer low prices. Hence, a mobile crowdsourcing scheme has been proposed [83] for drivers to observe the fuel price at different gas stations. As a user's car approaches a gas station, the user's mobile camera placed at the front of the car is activated automatically to capture photos of the price boards. The photos are then processed on the server, and the fuel price characters are identified through a neural network system.

Since there might be several routes to a destination, a smart navigator [84] has been designed to offer a personalized and optimal route based on the quality of road surface and the risk profile of the driver in addition to considering both time and distance. The surface quality of roads is obtained from the onboard diagnostic II (OBDII) unit and inertial sensors (*e.g.* accelerometer and gyroscope) of vehicles or smart devices of the drivers. These sensors measure the information of linear accelerations and angular rotations of a vehicle. The resultant information is then sent to the driver's smartphone via Bluetooth. In local processing, the information is first de-noised through wavelet packet decomposition (WPD), and the frequency of road anomalies is then obtained. The feature extraction (statistical, time, frequency, and time-frequency) and classification techniques are employed to identify and classify these anomalies. Ultimately, the information of road anomalies that are location stamped is sent to the server. In addition to inertial sensors, front and rear cameras are used along with short-range and long-range radar sensors to identify the driver's behavior. A hidden Markov model and the Viterbi algorithm are used to determine the driver's behavior. The identified behavior is then location stamped and sent to the server. The quality of a road is lastly classified into good, moderate, and poor classes through fuzzy inference. Moreover, the driver's behavior risk is determined at different segments of a route with respect to the identified behavior and environmental features (weather conditions). At the time of a route request, the navigator analyzes all the connected routes between the origin and the destination to find the optimal route. The information of different segments of routes will then be extracted from the database to obtain the overall quality level of a route based on the mean quality of its segments. In addition, the driver's profile risk history on similar segments of the road is extracted from the database. Based on the weighted mean, the overall risk of a road is obtained for the driver. It includes risky, moderate, and safe tags. Ultimately, the navigator can recommend the most optimal route to the driver based on the resultant information.

Playing appropriate music can prevent the driver's drowsiness and also affect the driver's mood. For this purpose, a crowdsourcing system [85] has been proposed for music selection while driving. Based on their feelings while listening to a group of songs, users label specific tags (*e.g.* lively, energetic, sad, groovy, noisy, and peaceful) along with their social context information (*e.g.* age, gender, and cultural background). Based on different social contexts, data is grouped and aggregated on the server. Ultimately, the appropriate song is played to drivers with respect to their human mood contexts.

## 5. Conclusion

This chapter reviewed and classified the smart city studies that adopted mobile crowdsourcing models. These systems have been classified into environmental, urban life, and transportation classes. In every reviewed system, one problem of smart cities has been discussed, and related solutions have been explained. Given the extensiveness of these studies, it can be concluded that mobile crowdsourcing has had a central role in solving the issues of smart cities. It is predicted that this powerful model will be used in different specialized areas of smart cities such as healthcare, archeology, tourism, social sciences, and confrontation with natural and unnatural disasters. Dealing with natural disasters (*e.g.* earthquakes, fires, floods, and viral epidemics) and unnatural ones (*e.g.* wars and terrorist attacks) are among the most serious challenges in smart cities. A common step in dealing with all of these disasters is to obtain accurate and dynamic information (in spatiotemporal dimensions) to make plans and manage responses. For this purpose, crowdsourcing can be used as a key strategy for collecting dynamic spatiotemporal information in future research studies.